\documentclass[twocolumn,runningheads]{svjour2}
\smartqed  % flush right qed marks, e.g. at end of proof
\usepackage{graphicx}
\journalname{Astrophysics and Space Science}
\begin{document}

\title{A Study of chiral property of field galaxies}
%\subtitle{}

\titlerunning{Chirality of field galaxies}        % if too long for running head

\author{B. Aryal \and R. Pandey \and W. Saurer}


\institute{First Author: B. Aryal \at
              $^{\rm 1}$Central Department of Physics, Tribhuvan University, Nepal\\
              \email{binil.aryal@uibk.ac.at}           %  \\
\and $^{\rm 1}$R. Pandey \at
              \email{panduram466@hotmail.com}
\and W. Saurer \at
              Institute of Astrophysics, Innsbruck University\\
              Technikstrasse 25/8, A-6020 Innsbruck, Austria\\
              \email{walter.saurer@uibk.ac.at}           %  \\
 }

\date{Received: 4 July 2013 / Accepted: ...........}

\maketitle

\begin{abstract}
We present an analysis of the chiral property of 1\,621 field
galaxies having radial velocity 3\,000 km s$^{-1}$ to 5\,000 km
s$^{-1}$. A correlation between the chiral symmetry breaking and
the preferred alignment of galaxies in the leading and trailing
structural modes is studied using chi-square, auto-correlation and
the Fourier tests. We noticed a good agreement between the random
alignment of the position angle (PA) distribution and the
existence of chirality in both the leading and trailing arm
galaxies. Chirality is found stronger for the late-type spirals
(Sc, Scd, Sd and Sm) than that of the early-types (Sa, Sab, Sb and
Sbc). A significant dominance (17\% $\pm$ 8.5\%) of trailing modes
is noticed in the barred spirals. In addition, chirality of field
galaxies is found to remain invariant under the global expansion.
The PA-distribution of the total trailing arm galaxies is found to
be random, whereas preferred alignment is noticed for the total
leading arm galaxies. It is found that the rotation axes of
leading arm galaxies tend to be oriented perpendicular the
equatorial plane. A random alignment is noticed in the
PA-distribution of leading and trailing spirals.


%\PACS{First \and Second \and More}
\end{abstract}

\section{Introduction}
\label{intro}

An object that is not superimposable on its mirror image is termed
as chiral object. Chiral objects do not show reflection symmetry,
but may exhibit rotational symmetry. They are not necessarily
asymmetric, because they can have other types of symmetry. An
object is achiral (non-chiral) if and only if it has an axis of
improper rotation (rotation by 360$^{\circ}$/$n$) followed by a
reflection in the plane perpendicular to this axis. Thus, an
object is chiral if and only if it lacks such an axis.

Chiral symmetry breaking or restoration is a key ingredient in
different problems of theoretical physics: from nonperturbative
quantum cromodynamics to highly doped semiconductors
(Garcia-Garcia \& Cuevas 2006). For a pseudoscalar meson, it is
found that the chiral symmetry breaking decreases with increasing
current-quark mass (Chang et al. 2007). Bagchi et al. (2006)
studied the large color approximation of the compact strange stars
and discussed the chiral symmetry restoration. They claimed that
the chiral symmetry restoration can be understood by exploring the
possible existence of strange stars in the Universe. In principle,
microscopic process have triggered huge astrophysical large scale
structure (Liddle \& Lyth 2000). It is therefore interesting to
study the breaking or restoration of chiral symmetry not only in
microscopic phenomena (L-neutrino, mesons, L-aminoacids, D-sugars,
etc), but also in macroscopic ones (stars, molecular clouds,
galaxies). According to the basic concepts of cosmology, initial
quantum fluctuations have been hugely enhanced during the
inflationary epoch, leading the formation of the large scale
structure in the Universe (Peacock 1999, Liddle \& Lyth 2000).
Thus, the macroscopic chirality should be related to some
primordial microscopic process which led to the today's observed
large scale structures (Fall 1982).

Differential rotation in a galaxy's disc generate density waves in
the disc, leading to spiral arms. According to gravitational
theory, the spiral arms born as leading and subsequently transform
to trailing modes. With the passage of time, the spiral pattern
deteriorates gradually by the differential rotation of the
equatorial plane of the galaxy, but the bar structure persists for
a long time (Oort 1970a). This structure can again regenerate
spiral pattern in the outer region. Thus, a close relation between
the origin of the arms in the spirals and barred spirals can not
be denied (Oort 1970b).

By considering the group of transformations acting on the
configuration space, Capozziello \& Lattanzi (2006) revealed that
the spiral galaxies possesses chiral symmetry in the large scale
structure. In addition, they predicted that the progressive loss
of chirality might have some connection with the
rotationally-supported (spirals, barred spirals) and randomized
stellar systems (lenticulars, ellipticals). Aryal, Acharya \&
Saurer (2007) carried out a study to test the Capozziello and
Lattanzi's (2006) prediction regarding the progressive loss of
chirality in the large scale structure. They analysed the
distribution of leading and trailing arm galaxies in the Local
Supercluster (LSC) and concluded the existence of chiral symmetry
for both the spirals and the barred spirals. However, the Virgo
cluster galaxies show a preferred alignment: the galactic rotation
axes of leading and trailing structures are found to lie in the
equatorial plane. Aryal \& Saurer (2005a) noticed a preferred
alignment for the late-type spirals and barred spirals in the
Local Supercluster. In addition, they found that the spin vector
projections of early- and late-type galaxies show opposite
alignment. Their results hint the existence of the chiral
characterization.

In this work, we present an analysis of chiral property of field
galaxies having radial velocity (RV) 3\,000 km s$^{-1}$ to 5\,000
km s$^{-1}$. We intend to study the importance of chiral symmetry
in order to understand the structural modes of the galaxy. In
addition, we expect to study the following: (1) Does chirality
exist for the spirals and barred spirals in the field? (2) Is
there any correlation between the chirality and preferred
alignment of galaxies? (3) Does the group of galaxies exhibit
chiral characterization? and finally, (4) What can we say about
the chiral and achiral (non-chiral) properties of the large scale
structure?

This paper is organized as follows: in Sect. 2 we describe the
method of data reduction. In Sect. 3 we give a brief account of
the methods and the statistics used. Finally, a discussion of the
statistical results and the conclusions are presented in Sects. 4
and 5.

\section{The sample: data reduction}

%--------------------------------------------------------------
% figure 1
\begin{figure}
\vspace{0.0cm}
      \centering
       \includegraphics[height=4.2cm]{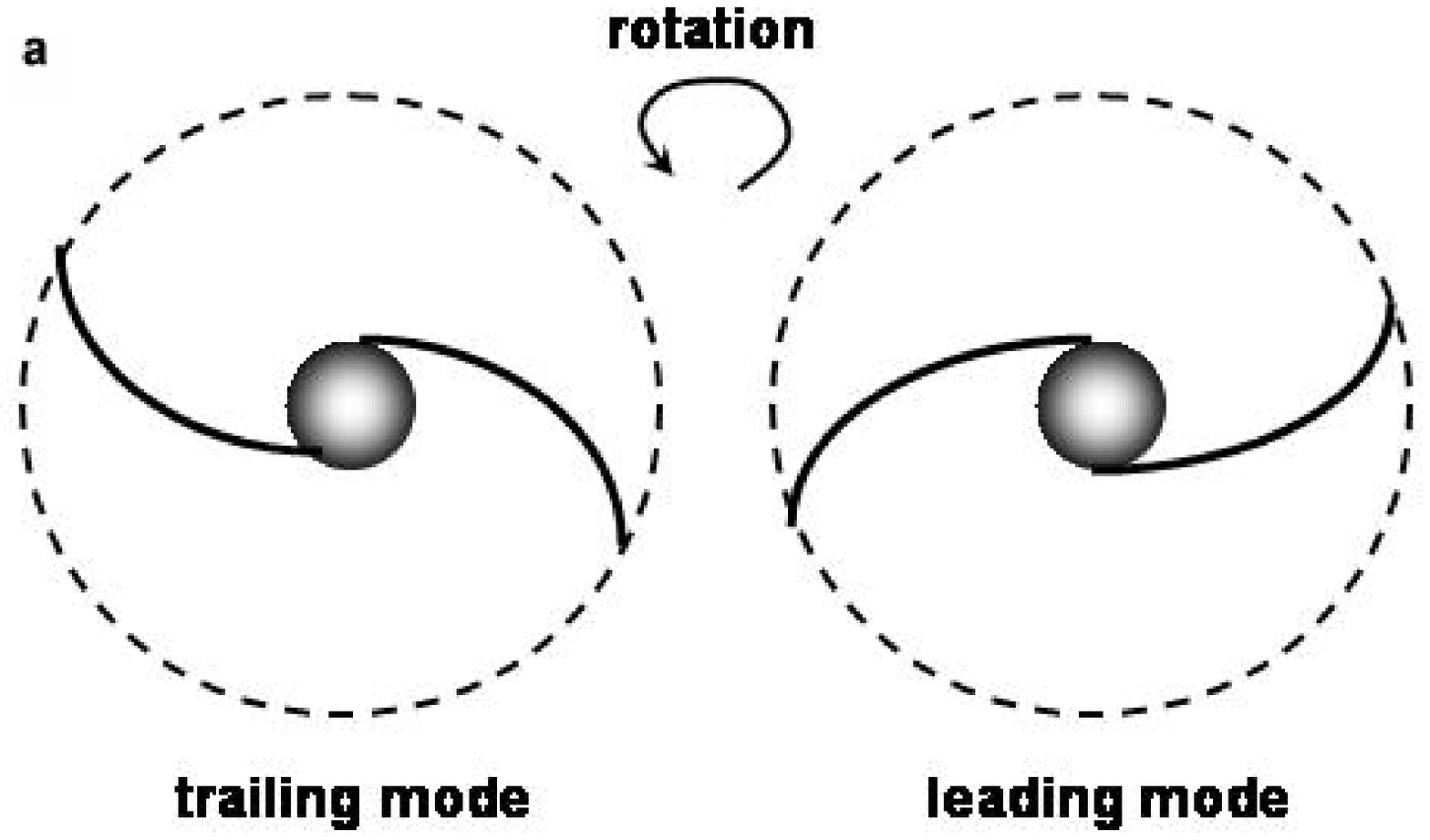}
       \includegraphics[height=4.5cm]{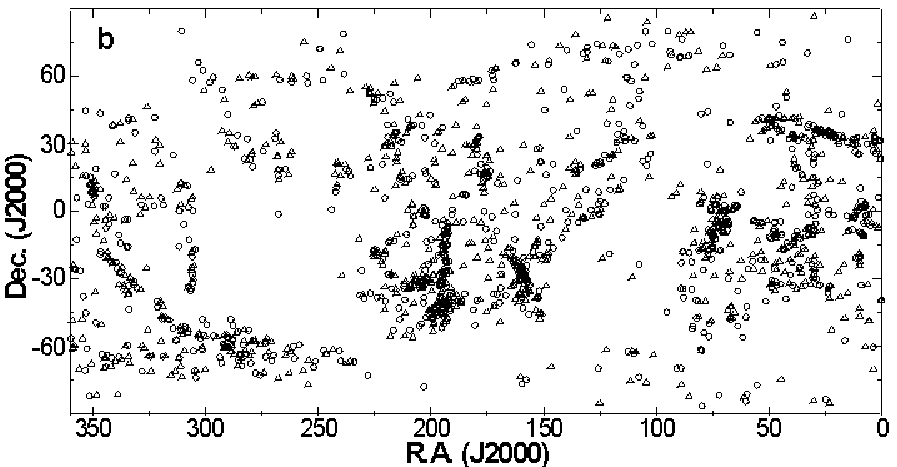}
       \includegraphics[height=4.1cm]{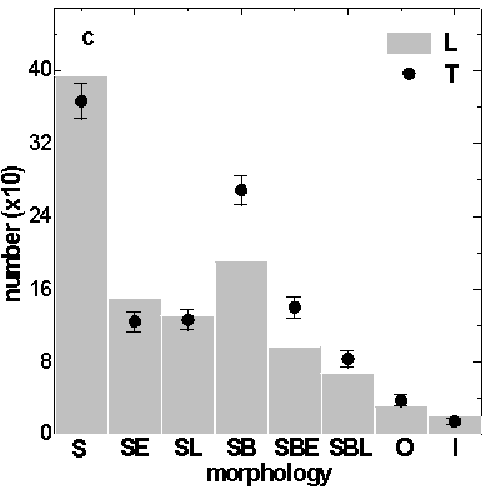}
       \includegraphics[height=4.1cm]{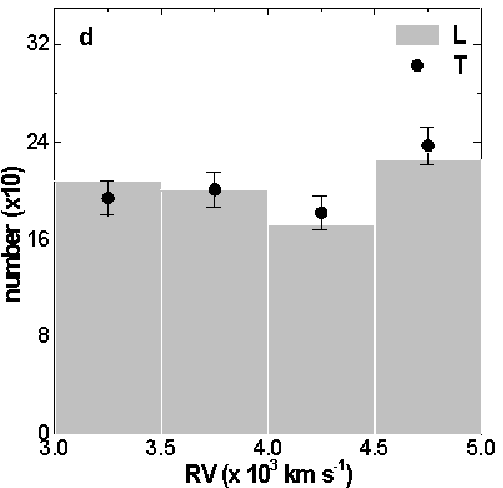}
       \includegraphics[height=4.1cm]{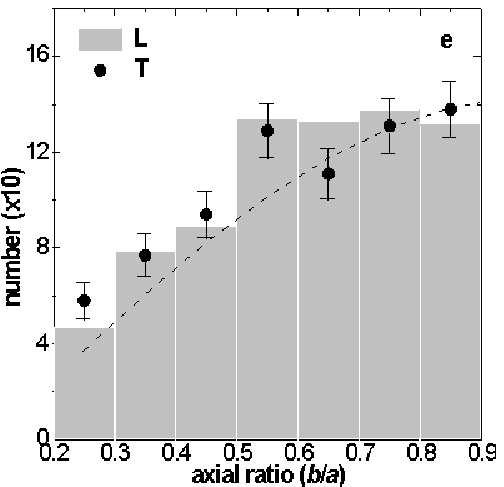}
       \includegraphics[height=4.1cm]{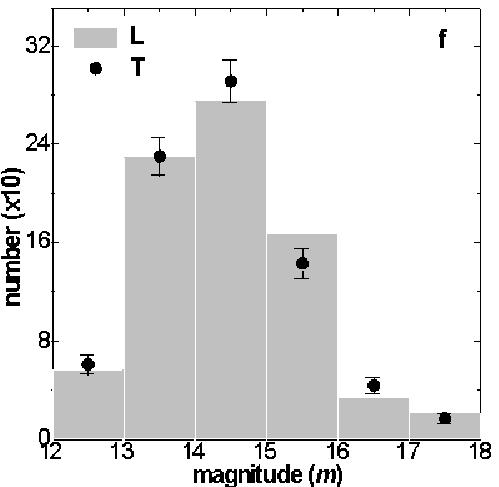}
      \caption[]{(a) A sketch representing the orientation (leading and trailing modes) of the spiral arms relative to
      the direction of rotation of the galaxy. (b) All-sky distribution
      of 1\,621 leading ($\triangle$) and trailing ($\circ$) arm field galaxies having
      RVs in the range 3\,000 km s$^{-1}$ to 5\,000 km s$^{-1}$.
      The morphology (c), radial velocity (d), axial ratio (e) and the
      magnitude (f) distribution of leading and trailing arm galaxies in our databse.
      The statistical $\pm$1$\sigma$ error bars are shown for the trailing ($\bullet$) subsample.
      The dashed line (e) represents the expected distribution. }
\end{figure}
%--------------------------------------------------------------

The method of data compilation and the identification of the
structural modes of the galaxies were the same as in Aryal,
Acharya \& Saurer (2007). We describe the selection criteria and
the data reduction process in brief here. A galaxy had to fulfill
the following selection criteria in order to be selected: (1) the
RV should lie in the range 3\,000 km s$^{-1}$ to 5\,000 km
s$^{-1}$, (2) the morphology should be known, (3) should not be
the cluster galaxy, (4) the diameters, magnitude and the position
angle should be given and, (5) the arms should be visible.

There were two clusters Abell 0426 ($\alpha$(J2000) = 03$^{\rm
h}$18$^{\rm m}$ 36.4$^{\rm s}$, $\delta$(J2000) =
+41$^{\circ}$30'54'') and Abell 3627 ($\alpha$(J2000) = 16$^{\rm
h}$15$^{\rm m}$32.8$^{\rm s}$, $\delta$(J2000) =
--60$^{\circ}$54'30'') in our region. These clusters have mean RVs
5\,366 km s$^{-1}$ (75 $\pm$ 5 Mpc) and 4\,881 km s$^{-1}$ (63
$\pm$ 4 Mpc), respectively (Abell, Corwin \& Olowin 1989, Struble
\& Rodd 1999). We removed the galaxies belong to the cluster Abell
0426 using the catalog established by Brunzendorf \& Meusinger
(1999). For the cluster Abell 3627 galaxies, we used Photometric
Atlas of Northern Bright Galaxies (Kodaira, Okamura, \& Ichikawa
1990) and Uppsala Galaxy Catalogue (Nilson 1973). There were 174
galaxies belongs to these clusters in our database.

The RVs were compiled from Las Campanas Redshift Survey (Shectman
1996). The PAs and the diameters of galaxies were added from the
Uppsala Galaxy Catalogue (Nilson 1973), Uppsala obs. General
Catalogue, Addendum (Nilson 1974), Photometric Atlas of Northern
Bright Galaxies (Kodaira, Okamura, \& Ichikawa 1990), ESO/Uppsala
Survey of the European Southern Observatory (Lauberts 1982),
Southern Galaxy Catalogue (Corwin et al. 1985) and Third Reference
Catalogue of Bright Galaxies (de Vaucouleurs et al. 1991).

In the NED (NASA/IPAC extragalactic database, http:
//nedwww.ipac.caltech.edu/), 6\,493 galaxies having RVs
3\,000 km s$^{-1}$ to 5\,000 km s$^{-1}$ were listed until the
cutoff date (October 2006). The diameters were given for 5\,324
(82\%) galaxies. Both the diameters and PAs were listed for 3\,971
(61\%) galaxies. Morphological information was given in the
catalogues for 3\,276 (50\%) galaxies. We inspected the DSS images
of 3\,276 galaxies using NED and ALADIN2.5 softwares. There were
2\,107 (32\%) galaxies having distinct spiral arms in our
database.

In order to understand true structural modes (leading or trailing)
of spiral galaxies, we need to know the direction of the spiral
pattern (S- or Z-shaped), the approaching and receding sides and
the near and far parts, since galaxies are commonly inclined in
space to the line of the sight. The S and Z-patterns can be
determined from the image of the galaxy. Similarly, the
approaching and receding sides can be defined if spectroscopy data
on rotation is available. The third one is fairly hard to
established. For this, Pasha (1985) used `tilt' criteria and
studied the sense of winding of the arms in 132 spirals. He found
107 spirals to have trailing arm. It should be remembered that the
classical `tilt' criteria is based on the visible asymmetry of a
dust matter distribution. Now it is well known that the dark
matter halo dominates the dynamics of the galaxies throughout. In
our own galaxy, the observed rotation of the stars and gas clouds
indicates that the visible matter is surrounded by a halo of this
dark matter containing the major portion of the total galaxy mass
and extending very far beyond the visible matter. The nature of
dark matter in the galactic halo of spiral galaxies is still
undetermined. Thomasson et al. (1989) studied theoretically and
performed $N$-body simulations in order to understand the
formation of spiral structures in retrograde galaxy encounters.
Interestingly, they noticed the importance of halo mass. They
concluded that the spirals having halos with masses larger than
the disk mass exhibit leading pattern. Thus, the makeup of
galactic haloes is important to cosmology in order to understand
the evolution of galaxies.

Sugai \& Iye (1995) used statistics and studied the winding sense
of galaxies (S- and Z-shaped) in 9\,825 spirals. No significant
dominance from a random distribution is noticed. Aryal \& Saurer
(2005a) studied the spatial orientations of spin vectors of 4\,073
galaxies in the Local Supercluster. No preferred alignment is
noticed for the total sample. These results hint that the
distribution of angular momentum of galaxies is entirely random in
two- (S- and Z-shaped) and three-dimensional (spin vector)
analysis provided the database is rich. Thus, choice of rotation
(approaching or receding, near or far parts) of individual galaxy
might be random for a observer. We used computer and gave random
direction to our galaxies. Each galaxy got a virtual direction by
the computer. After then, the arm patterns (S- or Z-type) the
galaxies were studied visually by one of the authors (PL) in order
to maintain homogeneity. The Digitized Sky Survey (DSS) image and
the contour maps of the galaxies were studied in order to identify
their structural modes. For this, we use MIDAS software. The
leading mode is one whose outer tip points towards the direction
of galactic rotation (see Fig. 1a). Similarly, the outer tip of
trailing mode directs in the direction opposite to the galactic
rotation. The re-examination of the arms using IRAF software
resulted the rejection of more than 23\% of the objects. These
rejected galaxies were nearly edge-on galaxies. As expected, it
was relatively easier to identify the structural modes of nearly
face-on than that of nearly edge-on galaxies.

In this way, we compiled a database of 1\,621 galaxies showing
either leading or trailing structural mode. {\bf For individual
galaxies, these structural modes might be incorrect. As a gross, it
might reveal the results in the subjective sense.} There were 807
leading and 821 trailing patterns in our database. All sky
distribution of leading and trailing arm galaxies is shown in Fig.
1b. The symbols ``$\circ$" and ``$\triangle$" represent the
positions of the trailing and the leading arm galaxies,
respectively. Several groups and aggregations of the galaxies can be
seen in the figure. The inhomogeneous distribution of the positions
of the galaxies might be the selection effects for the galaxy
orientation study (Aryal \& Saurer 2000).

The morphology, radial velocity, axial ratio and the magnitude
distributions of leading and the trailing arm field galaxies are
shown in Figs. 1c,d,e,f. The spirals (47\%) dominate our database
in the morphological distribution (Fig. 1c). However, a
significant dominance of trailing modes are noticed in the barred
spirals whereas a weak dominance of leading modes are found in the
spirals. The galaxies in the RV distribution ($\Delta$RV = 5\,00
km s$^{-1}$) were nearly equal (Fig. 1d). The axial ratio
distribution shows a good agreement with the expected {\it cosine}
curve in the limit 0.2 $<$ $b$/$a$ $\leq$ 0.9 (Fig. 1e). The
absolute magnitude lie between 13 and 16 for 82\% galaxies in our
database (Fig. 1f).

We classified the database into 34 subsamples for both the leading
and trailing modes on the basis of the morphology, radial
velocity, area and the group of the galaxies. A statistical study
of these subsamples are given in Table 1 and discussed in Sect.
3.1.

%__________________________________________________________________
\section{Method of analysis}
%__________________________________________________________________

Basic statistics is used to study the dominance of leading and
trailing modes. At first, morphology and RV dependence is studied
concerning the chiral property of galaxies. Secondly, sky is divided
into 16 equal parts in order to observe the violation of chirality
locally. Several galaxy groups are identified in the all-sky map
where the structural dominance are noticed. Finally, we study the
dominance of leading or trailing arm galaxies in these groups.

We assume isotropic distribution as a theoretical reference and
studied the equatorial PA-distribution in the total sample and
subsamples. In order to measure the deviation from isotropic
distribution we have carried out three statistical tests:
chi-square, auto-correlation and the Fourier.

We set the chi-square probability P($>\chi^2$) = 0.050 as the
critical value to discriminate isotropy from anisotropy, this
corresponds to a deviation from isotropy at the 2$\sigma$ level
(Godlowski 1993). Auto correlation test takes account the
correlation between the number of galaxies in adjoining angular
bins. We expect, auto correlation coefficient C$\rightarrow$0 for
an isotropic distribution. The critical limit is the standard
deviation of the correlation coefficient C.

If the deviation from isotropy is only slowly varying with angles
(in our case: PA) the Fourier test can be applied.

A method of expanding a function by expressing it as an infinite
series of periodic functions ({\it sine} and {\it cosine}) is
called Fourier series. Let $N$ denote the total number of
solutions for galaxies in the sample, $N$$_{k}$ the number of
solutions in the k$^{th}$ bin, $N$$_{0}$ the mean number of
solutions per bin, and $N$$_{0k}$ the expected number of solutions
in the k$^{th}$ bin. Then the Fourier series is given by (taking
first order Fourier mode),

%_________________________________________________
\begin{equation}
\begin{array}{l}
N_{k} = N_{k}(1+ \Delta_{11} \cos 2\beta_{k}+ \Delta_{21} \sin 2\beta_{k}+ ......)\\
\end{array}
\end{equation}
%_________________________________________________
Here the angle {\bf $\beta$$_{k}$} represents the polar angle in
the k$^{th}$ bin. The Fourier coefficients  $\Delta_{11}$ and
$\Delta_{21}$ are the parameters of the distributions. We obtain
the following expressions for the Fourier coefficients
$\Delta_{11}$ and $\Delta_{21}$,
%_________________________________________________
\begin{equation}
\begin{array}{l}
\Delta_{11} = \sum (N_{k}-N_{0k}) \cos 2\beta_{k} / \sum N_{0k} \cos^2 2\beta_{k} \\
\end{array}
\end{equation}
%_________________________________________________
\begin{equation}
\begin{array}{l}
\Delta_{21} = \sum (N_{k}-N_{0k}) \sin 2\beta_{k} / \sum N_{0k} \sin^2 2\beta_{k} \\
\end{array}
\end{equation}
%________________________________________________
The standard deviations  ($\sigma$($\Delta_{11}$)) and
($\sigma$($\Delta_{21}$)) can be estimated using the expressions,
%________________________________________________
\begin{equation}
\begin{array}{l}
\sigma (\Delta_{11}) = (\sum N_{0k} \cos^2 2\beta_{k})^{-1/2} \\
\end{array}
\end{equation}
%_________________________________________________
\begin{equation}
\begin{array}{l}
\sigma (\Delta_{21}) = (\sum N_{0k} \sin^2 2\beta_{k})^{-1/2} \\
\end{array}
\end{equation}
%__________________________________________________
The probability that the amplitude
\begin{equation}
\begin{array}{l}
\Delta_{1} = (\Delta_{11}^2 + \Delta_{21}^2)^{1/2} \\
\end{array}
\end{equation}
%__________________________________________________
greater than a certain chosen value is given by the formula
\begin{equation}
\begin{array}{l}
P(>\Delta_{1}) = \exp(-nN_{0}\Delta_{1}^2/4) \\
\end{array}
\end{equation}
%_________________________________________________
with standard deviation
\begin{equation}
\begin{array}{l}
\sigma (\Delta_{1}) = (2/nN_{0})^{1/2} \\
\end{array}
\end{equation}
%___________________________________________________

The Fourier coefficient $\Delta_{11}$ gives the direction of
departure from isotropy. The first order Fourier probability
function $P$($>$$\Delta_{1}$) estimates whether (smaller value of
$P$($>$$\Delta_{1}$) or not (higher value of $P$($>$$\Delta_{1}$)
a pronounced preferred orientation occurs in the sample.

%________________________________________________________
\section{Results}

First we present the statistical result concerning the
distribution of leading and trailing modes of galaxies in the
total sample and subsamples. Then, we study the distribution of
the leading and trailing arm galaxies in the unit area of the sky
and the groups. The equatorial PA-distribution of galaxies in the
total sample and subsamples is presented. At the end, a general
discussion and a comparison with the previous results will be
presented.

%________________________________________________________
\subsection{Distribution of leading and trailing structures}

A statistical comparison between the total sample and subsamples
of the leading and trailing arm field galaxies is given in Table
1. Fig. 2 shows this comparison in the histogram. The $\Delta$(\%)
in Table 1 and Fig. 2 represent the percentage difference between
the number of trailing and leading arm galaxies. We studied the
standard deviation ($sde$) of the major diameters ($a$) of
galaxies in the total sample and subsamples for both the leading
and trailing modes. In Table 1, $\Delta(a\,sde)$ represents the
difference between the standard deviation of the major diameters
of leading and the trailing arm galaxies.

An insignificant difference (0.4\% $\pm$ 0.2\%) between the total
number of trailing and the leading arm galaxies are found (Table
1). The difference between the standard deviation of the major
diameters ($\Delta(a\,sde)$) of the trailing and leading arm
galaxies is found less than 0.019 (seventh column, Table 1).
Interestingly, the sum of the major diameters of total trailing
and leading arm galaxies coincide. This result strongly suggests
the existence of chirality of field galaxies having RV 3\,000 km
s$^{-1}$ to 4\,000 km s$^{-1}$.

In Fig. 2, the slanting-line (grey-shaded) region corrosponds to
the region showing $\leq$ 10\% (5\%) $\Delta$ value. Almost all
subsamples lie within this region, suggesting the existence of
chirality within 10\% error limit. We present the distribution of
the subsamples of leading and trailing arm galaxies classified
according as their morphology, RVs, area and the groups below.

\subsubsection{Morphology}

In the spirals, leading structural modes are found 3.7\% $\pm$
1.8\% more than that of trailing modes. The chirality is found
stronger for the late-type spirals (Sc, Scd, Sd and Sm) than that
of early-type (Sa, Sab, Sb and Sbc): $\Delta$ value turned out to
be 9.5\% ($\pm$ 4.8\%) and 1.8\% ($\pm$ 1.0\%) for early- and
late-types (Table 1). Thus, the late-type spirals are the best
candidate of the chiral object in our database.

%--------------------------------------------------------------
% figure 1
\begin{figure}
\vspace{0.0cm}
      \centering
      \includegraphics[height=4.4cm]{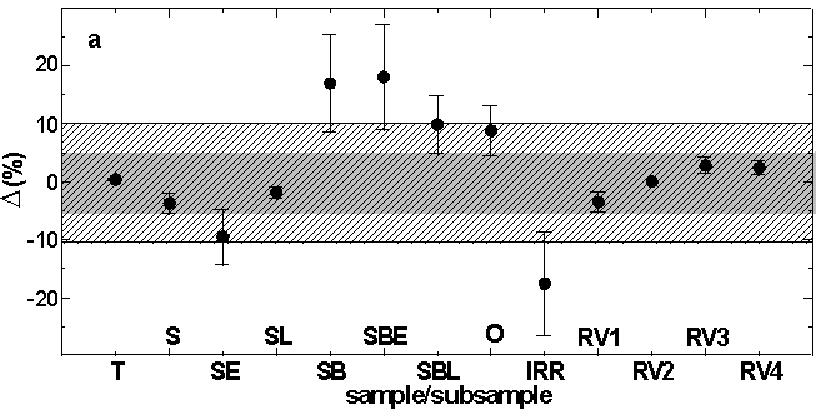}
      \includegraphics[height=4.4cm]{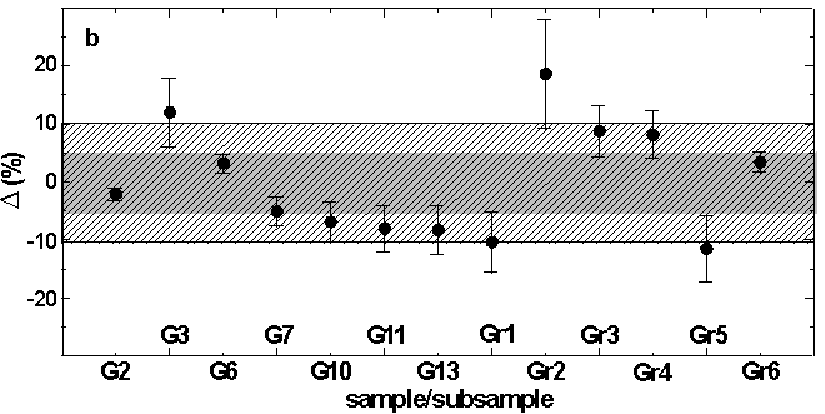}
      \caption[]{The basic statistics of the leading and trailing
      arm galaxies in the total sample and subsamples. The full form of the abbreviations (X-axis)
      are given in Table 1 (first column). $\Delta$(\%)=$T$-$L$, where $T$
      and $L$ represent the number of trailing and leading arm galaxies, respectively.
      The statistical error bars $\sigma$(\%) shown in the figure are calculated as: $\sigma$(\%) =
$\sigma$/($\sqrt{T}$+$\sqrt{L}$)$\times$100, where $\sigma$ =
($\sqrt{T}$-$\sqrt{L}$). The grey-shaded and the slanting-line
region represent the $\leq$$\pm$5\% and $\leq$$\pm$10\% $\Delta$
value, respectively. }
\end{figure}
%--------------------------------------------------------------

%-------------------------------------------------------------------
% table 2
\begin{table*}
      \caption[]{Statistics of leading (column 3) and trailing arm (column 4)
      galaxies in the total sample and subsamples. The fifth and sixth column give the numeral and percentage
      difference ($\Delta$ = $T$--$L$) between the trailing ($T$) and the leading ($L$) structural modes.
      The next two columns give the error: $\sigma$ = ($\sqrt{T}$--$\sqrt{L}$) and
      $\sigma$(\%) = $\sigma$/($\sqrt{L}$+$\sqrt{T}$)$\times$100. The eighth column gives the difference between the
      standard deviation (in arcmin) of the major diameters ($a$) of the trailing and leading arm galaxies ($\Delta$($a\,sde)$).
      The difference between the sum of the major diameters ($\Delta$($a$)\%) are listed in the last column.
      The sample/subsample and their abbreviations are given in first two columns.}
    $$
\begin{array}{p{0.23\linewidth}rccrrrrrr}
\hline\hline \noalign{\smallskip}
sample/subsample & $symbol$  & $L$ & $T$ & $$\Delta$$ & $$\Delta$(\%)$ &  $$\sigma$(\%)$ & $$\Delta$($a$\,sde)$ & $$\Delta$($a$)(\%)$\\

\hline\hline \noalign{\smallskip}
Total                                 &   $T$     &   814 &   807 &   7   &   0.4     &   0.2  &  0.019 &  0.0 \\
Spiral                                &   $S$     &   395 &   367 &   -28 &   -3.7    &   -1.8 &  0.031 &  3.0 \\
Spiral (early-type)                   &   $SE$    &   150 &   124 &   -26 &   -9.5    &   -4.8 &  0.058 &  8.1 \\
Spiral (late-type)                    &   $SL$    &   131 &   126 &   -5  &   -1.9    &   -1.0 &  0.031 &  0.3 \\
Barred Spiral                         &   $SB$    &   191 &   269 &   78  &   17.0    &   8.5  &  0.062 &  15.2 \\
Barred Spiral (early-type)            &   $SBE$   &   97  &   140 &   43  &   18.1    &   9.1  &  0.091 &  14.6 \\
Barred Spiral (late-type)             &   $SBL$   &   68  &   83  &   15  &   9.9     &   5.0  &  0.066 &  9.2 \\
Unknown\,Morphology                   &   $O$     &   31  &   37  &   6   &   8.8     &   4.4  &  0.095 &  3.9 \\
Irregular                             &   $I$     &   20  &   14  &   -6  &   -17.6   &   -8.9 &  0.105 &  5.2 \\
3\,000$<$RV (km s$^{-1}$)$\leq$3\,500 &   $RV1$   &   194 &   208 &   -14 &   -3.5    &   -1.7 &  0.046 &  2.6 \\
3\,500$<$RV (km s$^{-1}$)$\leq$4\,000 &   $RV2$   &   201 &   201 &   0   &   0.0     &   0.0  &  0.031 &  3.1 \\
4\,000$<$RV (km s$^{-1}$)$\leq$4\,500 &   $RV3$   &   182 &   172 &   10  &   2.8     &   1.4  &  0.034 &  0.6 \\
4\,500$<$RV (km s$^{-1}$)$\leq$5\,000 &   $RV4$   &   237 &   226 &   11  &   2.4     &   1.2  &  0.041 &  1.0 \\
Grid 1                                &   $G1$    &   21  &   20  &   1   &   2.4     &   1.2   & 0.068 &  6.8 \\
Grid 2                                &   $G2$    &   116 &   121 &   -5  &   -2.1    &   -1.1  & 0.014 &  1.6 \\
Grid 3                                &   $G3$    &   112 &   88  &   24  &   12.0    &   6.0   & 0.076 &  9.3 \\
Grid 4                                &   $G4$    &   14  &   12  &   2   &   7.7     &   3.9   & 0.647 &  9.7 \\
Grid 5                                &   $G5$    &   8   &   11  &   -3  &   -15.8   &   -7.9  & 0.042 &  22.3 \\
Grid 6                                &   $G6$    &   80  &   75  &   5   &   3.2     &   1.6   & 0.081 &  4.5 \\
Grid 7                                &   $G7$    &   56  &   62  &   -6  &   -5.1    &   -2.5  & 0.095 &  8.9 \\
Grid 8                                &   $G8$    &   33  &   22  &   11  &   20.0    &   10.1  & 0.073 &  12.9 \\
Grid 9                                &   $G9$    &   31  &   20  &   11  &   21.6    &   10.9  & 0.028 &  18.1 \\
Grid 10                               &   $G10$   &   108 &   124 &   -16 &   -6.9    &   -3.5  & 0.004 &  5.4 \\
Grid 11                               &   $G11$   &   52  &   61  &   -9  &   -8.0    &   -4.0  & 0.025 &  6.2 \\
Grid 12                               &   $G12$   &   20  &   20  &   0   &   0.0     &   0.0   & 0.409 &  7.4 \\
Grid 13                               &   $G13$   &   66  &   78  &   -12 &   -8.3    &   -4.2  & 0.039 &  2.9 \\
Grid 14                               &   $G14$   &   44  &   37  &   7   &   8.6     &   4.3   & 0.356 &  10.3 \\
Grid 15                               &   $G15$   &   44  &   47  &   -3  &   -3.3    &   -1.6  & 0.050 &  5.2 \\
Grid 16                               &   $G16$   &   9   &   9   &   0   &   0.0     &   0.0   & 0.191 &  8.6 \\
Group 1                               &   $Gr1$   &   30  &   37  &   -7  &   -10.4   &   -5.2  & 0.032 &  7.1 \\
Group 2                               &   $Gr2$   &   70  &   48  &   22  &   18.6    &   9.4   & 0.097 &  12.6 \\
Group 3                               &   $Gr3$   &   37  &   31  &   6   &   8.8     &   4.4   & 0.027 &  4.3 \\
Group 4                               &   $Gr4$   &   40  &   34  &   6   &   8.1     &   4.1   & 0.031 &  3.9 \\
Group 5                               &   $Gr5$   &   85  &   107 &   -22 &   -11.5   &   -5.7  & 0.089 &  11.2 \\
Group 6                               &   $Gr6$   &   45  &   42  &   3   &   3.4     &   1.7   & 0.024 &  1.6 \\
\noalign{\smallskip} \hline\hline
\end{array}
     $$
\end{table*}
%------------------------------------------------------------------

The dominance of trailing structural modes are significant (17\%
$\pm$ 8.5\%) in spiral barred galaxies. The $\Delta$ value is
found $>$ 9\% for both early- (SBa, SBab, SBb and SBbc) and
late-type (SBc, SBcd, SBd, SBm) barred spirals. Thus, we suspect
that the field SB galaxies are not the best candidates of chiral
objects. Similar result (i.e., $\Delta$ $>$ 8\%) is found for the
irregulars and the morphologically unidentified galaxies.

One interesting similarity is noticed between the late-type
spirals and barred spirals. The $\Delta$ value for both the
late-types are found less than that of early-types (see Table 1).
Thus, the chirality is favourable for the late-types rather than
the early-types.

The difference between the standard deviation of the major
diameters ($\Delta(a\,sde)$) for trailing and leading arm galaxies
is found less than 0.050 arc minute for the total sample, spirals
and the late-type spirals (seventh column, Table 1). These samples
showed $\Delta$ value $<$ 5\% (grey-shaded region, Fig. 2a). Thus,
we found a good correlation between the $\Delta$(\%) and
$\Delta(a\,sde)$ value. Probably, this result hints the fact that
the size of the non-superimposable mirror images should lie within
a limit. In our database, this limit should not exceed 0.050 arc
minute for $\Delta(a\,sde)$.

The difference between the sum of the major diameters (in
percentage) are found greater than 10\% for the barred spirals and
early-type barred spirals. Interestingly, these two subsamples
showed $\Delta$ value greater than 15\% (Fig. 2a).  We suspect
that the SB galaxies possess chiral symmetry breaking.

Thus, we found that the chirality between the total leading and
trailing arm galaxies exist in our database. This behavior is
found prominent for the spirals, mainly for early-type spirals.

\subsubsection{Radial velocity}

A very good correlation between the number of leading and trailing
arm galaxies can be seen in the RV classifications (Fig. 1d). All
4 subsamples show the $\Delta$ and $\Delta(a\,sde)$ value less
than 5\% and 0.050, respectively (Table 1). In addition,
$\Delta$($a$) is found to be $<$ 5\%. This result is important in
the sense that the statistics in these subsamples is rich (number
of galaxies $>$ 170) enough. Thus, we could not observe the
violation of chirality in the low and high RV galaxies in our
database.

A difference is noticed: dominance of leading and trailing modes
in low (RV1) and high (RV3, RV4) RV subsamples, respectively.
However, this dominance is not significant (i.e., $\Delta$ $<$
5\%). An equal number of leading and trailing arm galaxies are
found in the subsample RV2 (3\,500 $<$ RV (km s$^{-1}$) $\leq$
4\,000) (Table 1). This might be a coincidence. In order to check
the binning effect, we further classify the total galaxies in 6
($\Delta$RV = 333 km s$^{-1}$) and 8 bins ($\Delta$RV = 250 km
s$^{-1}$) and study the statistics. No significant dominance of
either trailing or leading structural modes are noticed.

Thus, it is found that the chirality of field galaxies remain
invariant with the global expansion (i.e, expansion of the
Universe). This is an important result. We further discuss this
result below.

\subsubsection{Area}

We study the distribution of leading and trailing arm galaxies by
dividing the sky into 16 equal parts (Fig. 3a). The area of the
grid (G) is 90$^{\circ}$ $\times$ 45$^{\circ}$ (RA $\times$ Dec).
The area distribution of leading and trailing arm galaxies are
plotted, that can be seen in Fig. 3a'. The statistical parameters
are given in Table 1.

%--------------------------------------------------------------
% figure 1
\begin{figure}
\vspace{0.0cm}
      \centering
      \includegraphics[height=4.6cm]{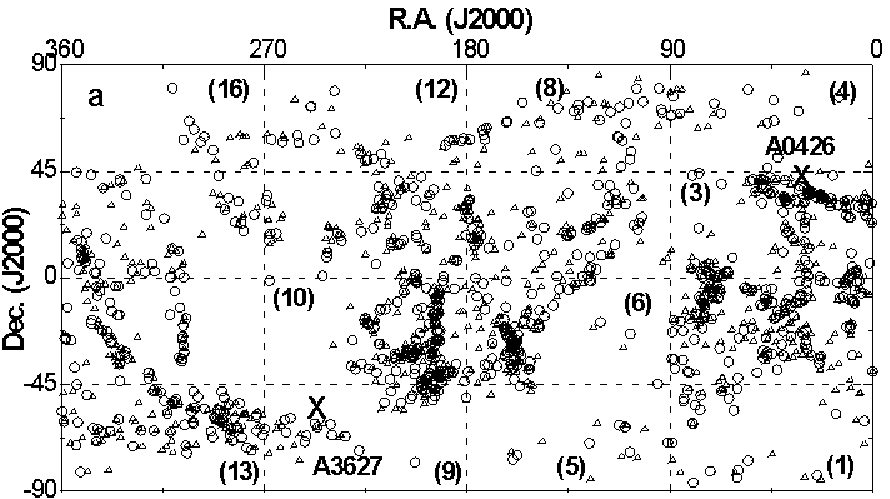}
      \includegraphics[height=3.0cm]{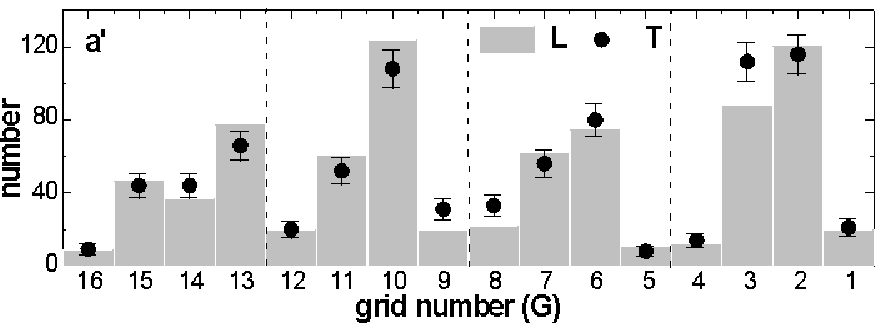}
      \includegraphics[height=4.6cm]{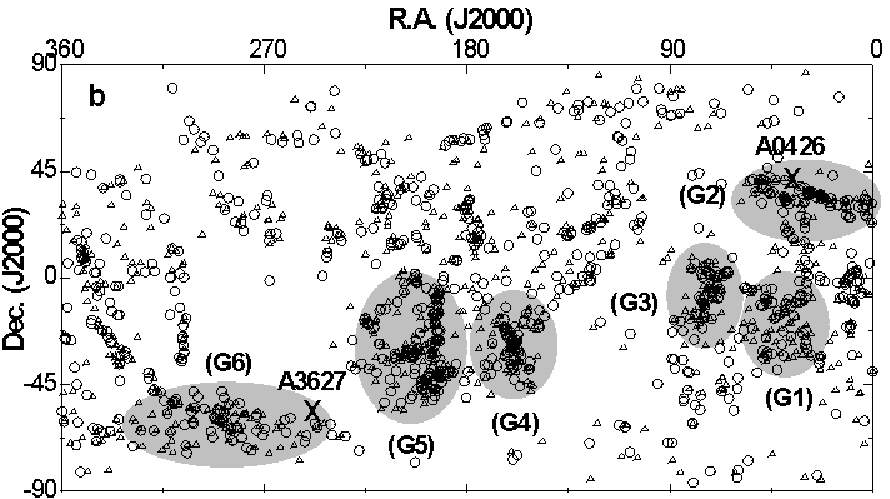}
      \includegraphics[height=3.0cm]{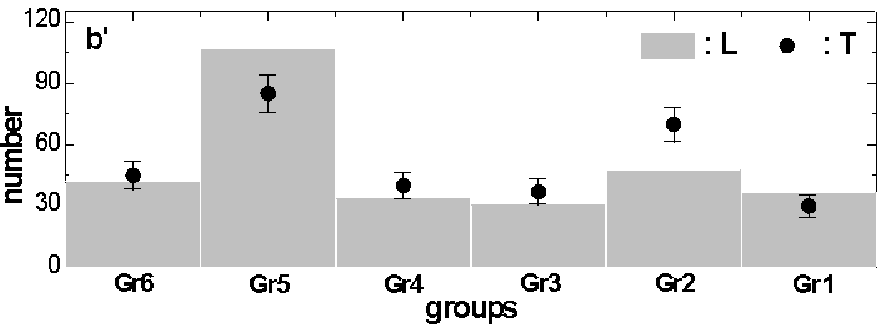}
      \caption[]{(a) All sky distribution of leading (hollow circle) and trailing (hollow triangle)
      arm galaxies in 16 area grids. (a') The histogram showing the
      distribution of the leading and trailing arm galaxies in the grids G1 to G16.
      (b) Six groups of the galaxies, represented by the grey-shaded region.
      (b') The distribution of leading and trailing structures in
      6 groups. The statistical error bar $\pm$1$\sigma$ is shown. The
positions of the clusters Abell 0426 and Abell 3627 are shown by
the symbol ``$\times$" (a,b).}
\end{figure}
%--------------------------------------------------------------

A significant dominance ($>$2$\sigma$) of trailing structures is
noticed in grid 3 (RA: 0$^{\circ}$ to 90$^{\circ}$, Dec:
0$^{\circ}$ to 45$^{\circ}$ (J2000)) (Fig. 3a,a'). An elongated
subcluster like structure can be seen in this grid. In this grid,
$\Delta$, $\Delta(a\,sde)$ and $\Delta(a)\%$ are found to be 12\%
$\pm$ 6\%, 0.076 and 9.3\%, respectively. These figures suggest
that the galaxies in G3 lost their chiral property. Probably, this
might be due to the apparent subgroupings or subclusterings of the
galaxies.

The trailing arm galaxies dominate in the grids G8 and G9 (Fig.
3a'). However, the statistics is poor ($<$ 40) in these grids
(Table 1). In addition, no groupings or subclustering are noticed.
Thus, we conclude nothing for these area grids.

A dominance ($\sim$1.5$\sigma$) of leading structures is noticed
in G10 (RA: 180$^{\circ}$ to 270$^{\circ}$, Dec: --45$^{\circ}$ to
0$^{\circ}$ (J2000)) and G13 (RA: 270$^{\circ}$ to 360$^{\circ}$,
Dec: --90$^{\circ}$ to --45$^{\circ}$ (J2000)) (Fig. 3a,a'). In
both the grids, a large aggregation of the galaxies can be seen. A
subcluster-like aggregation can be seen in G10. An elongated
structure can be seen in G13. In both the grids, $\Delta$ value is
found to be greater than 5\% (Table 1). Thus, we suspect that the
galaxies in these grids (G10, G13) are loosing chiral symmetry.
This result might reveal the effects of the cluster evolution on
chiral symmetry of galaxies.

No dominance of either leading or trailing structures is noticed
in the groups G1, G2, G4, G5, G6, G7, G11, G12, G14, G15 and G16.
Thus, the chirality is found intact in $\sim$ 80\% area of the
sky. We suspect that the groupings or subclusterings of the
galaxies lead the violation of chirality in the grids G3, G10 and
G13. We study the existence of chirality in the groups below.

\subsubsection{Galaxy groups}

In all-sky map, several groups of galaxies can be seen (Fig. 3a).
It is interesting to study the existence of chirality in these
groups. For this, we systematically searched for the groups
fulfilling following selection criteria: (a) major diameter $>$
30$^{\circ}$, (b) cutoff diameter $<$ 2 times the background
galaxies, (c) number of galaxies $>$ 50. We found 6 groups
fulfilling these criteria (Fig. 3b). All 6 groups (Gr) are
inspected carefully. In 3 groups (Gr2, Gr5 and Gr6), subgroups can
be seen. The number of galaxies in the groups Gr2 and Gr5 are
found more than 100.

The clusters Abell 0426 and Abell 3627 are located close to the
groups Gr2 and Gr6. The symbol ``$\times$" represents the cluster
center in Fig. 3b. The mean radial velocities of these clusters
are 5\,366 km s$^{-1}$ and 4\,881 km s$^{-1}$, respectively.
However, we have removed the member galaxies of these clusters
from our database.

A significant dominance ($>$2$\sigma$) of trailing structures is
noticed in the group Gr2 (Fig. 3b,b'). The $\Delta$,
$\Delta(a\,sde)$ and $\Delta(a)\%$ values are found to be 18.6\%
$\pm$ 9.4\%, 0.097 and 12.6\%, respectively (Table 1). These
values indicate that the galaxies in this group might lost their
chiral symmetry. We suspect that the galaxies in this group is
under the influence of the cluster Abell 0426, due to which
apparent subclustering of the galaxies can be seen in this group.
This subclustering lead the violation of chiral symmetry.

The galaxies in Gr5 shows an opposite preference: a significant
dominance of the leading arm galaxies ($>$2$\sigma$) (Fig. 3b,b').
In this group, $\Delta$, $\Delta(a\,sde)$ and $\Delta(a)\%$ are
found to be 11.5\% $\pm$ 5.7\%, 0.089 and 11.2\%, suggesting the
violation of chirality (Table 1).

No humps or dips can be seen in the groups Gr1, Gr3, Gr4 and Gr6
(Fig. 3.2b,b'). Thus, the galaxies in these groups show chiral
property. It is interesting that the number of galaxies in these
groups are less than 100. The groups Gr2 and Gr5, which showed a
significant difference between the number of leading and trailing
arm galaxies, have a very good statistics (i.e., $>$ 100). Thus,
we conclude that the large aggregation of the galaxies lead the
violation of chirality.

In the group 6, we could not notice the influence of the cluster
Abell 3627. This might be due to the off location of the cluster
center from the group center.

%--------------------------------------------------------------
\subsection{Anisotropy in the position angle distribution}

%----------------------------------------------------------
% figure 8
\begin{figure} \vspace{0.0cm}
      \centering
      \includegraphics[height=4cm]{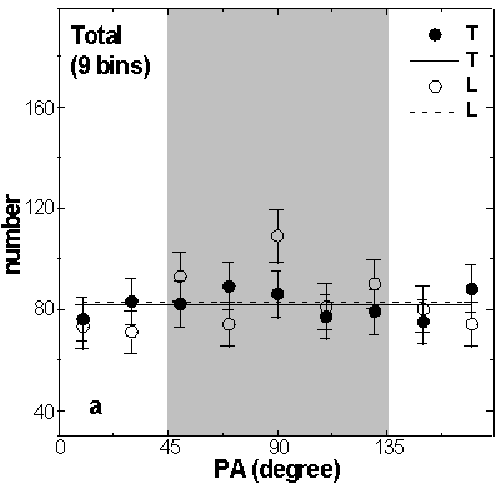}
      \includegraphics[height=4cm]{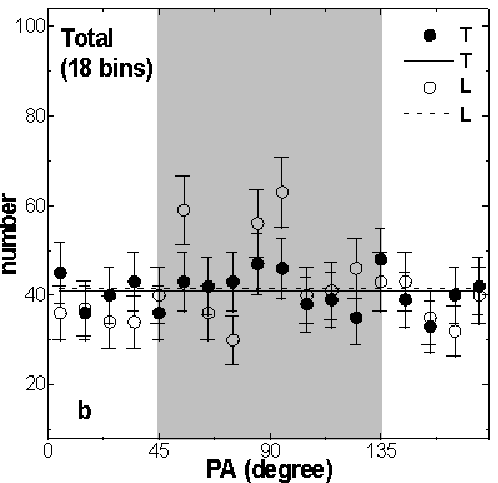}
      \caption[]{The equatorial position angle (PA) distribution of
      total leading and total trailing arm galaxies plotted in 9 (a) and 18 (b)
      bins. The solid and the dashed line represent the expected
      isotropic distribution for trailing and leading arm
      galaxies, respectively.
      The observed counts with statistical $\pm$1$\sigma$ error bars are shown. PA =
      90$^{\circ}$$\pm$45$^{\circ}$ (grey-shaded region)
      corresponds to the galactic rotation axes tend to be oriented perpendicular
      with respect to the equatorial plane.}
\end{figure}
%-----------------------------------------------------------

We study the equatorial position angle (PA) distribution of
trailing and leading arm galaxies in the total sample and the
subsamples. A spatially isotropic distribution is assumed in order
to examine non-random effects in the PA-distribution. In order to
discriminate the deviation from the randomness, we use three
statistical tests: chi-square, auto correlation and the Fourier.
The bin size was chosen to be 20$^{\circ}$ (9 bins) in all these
tests. The statistically poor bins (number of solution $<$ 5) are
omitted in the analysis. The conditions for anisotropy are the
following: the chi-square probability P($>\chi^2$) $<$ 0.050,
correlation coefficient $C$/$\sigma(C)$ $>$ 1, first order Fourier
coefficient $\Delta_{11}$/$\sigma(\Delta_{11}$) $>$ 1 and the
first order Fourier probability P($>\Delta_1$)$<$0.150 as used by
Godlowski (1993, 1994). Table 2 lists the statistical parameters
for the total samples and subsamples.

In the Fourier test, $\Delta_{11}$ $<$ 0 (i.e., negative)
indicates an excess of galaxies with the galactic plane parallel
to the equatorial plane. In other words, a negative $\Delta_{11}$
suggests that the rotation axis of galaxies tend to be oriented
perpendicular with respect to the equatorial plane. Because the
galactic plane is perpendicular to the rotation axis of the
galaxy. Similarly, $\Delta_{11}$ $>$ 0 (i.e., positive) indicates
that the rotation axis of galaxies tend to lie in the equatorial
plane.

In the histograms (see Figs. 4-7), a hump at
90$^{\circ}$$\pm$45$^{\circ}$ (grey-shaded region) suggests that
the galactic planes of galaxies tend to lie in the equatorial
plane. In other words, the rotation axes of galaxies tend to be
oriented perpendicular with respect to the equatorial plane when
there is excess number of solutions in the grey-shaded region in
the histograms.

All three statistical tests show isotropy in the total trailing
arm galaxies. Thus, no preferred alignment is noticed for the
total trailing arm field galaxies (solid circles in Fig. 4a).
Interestingly, all three statistical tests show anisotropy in the
total leading arm galaxies. The chi-square and Fourier
probabilities (P$(>\chi^2)$, P($>\Delta_1$)) are found 1.5\% ($<$
5\% limit) and 8.5\% ($<$ 15\% limit), respectively (Table 2). The
auto correlation coefficient (C/C($\sigma$)) turned --3.2 ($>>$1).
The $\Delta_{11}$/$\sigma(\Delta_{11}$) value is found to be
negative at $\sim$ 2$\sigma$ level, suggesting that the rotation
axes of leading arm galaxies tend to be oriented the equatorial
plane. Three humps at 50$^{\circ}$ ($>$1.5$\sigma$), 90$^{\circ}$
($>$2$\sigma$) and 130$^{\circ}$ (1.5$\sigma$) can be seen (Fig.
4a). All these humps lie in the grey-shaded region. We checked the
binning biasness in the statistics by increasing the number of
bins to 12 and 16. A similar statistical result is found for both
structural modes. Fig. 4b shows the PA-distribution histogram for
the total sample in 18 bins. The leading arm galaxies show three
significant humps in the grey-shaded region, supporting the above
mentioned result.

Thus, we conclude isotropy for trailing whereas anisotropy for
leading arm galaxies in the total sample.
%----------------------------------------------------------------
% table 3
\begin{table*}
      \caption[]{Statistics of the PA-distribution of galaxies in the
      total sample and subsamples (first column).
The second, third, fourth and fifth columns give the chi-square
probability (P$(>\chi^2)$), correlation coefficient
(C/C($\sigma$)), first order Fourier coefficient
($\Delta_{11}$/$\sigma$($\Delta_{11}$)), and first order Fourier
probability P($>\Delta_1$), respectively. The last four columns
repeats the previous columns.}
    $$
         \begin{array}{p{0.1\linewidth}ccccccccc}
            \hline
            \noalign{\smallskip}
            sample  &   & $Trailing$ &    &  &   & $Leading$ &  &   \\
            \noalign{\smallskip}
                    &  $P$(>\chi^2)$$ & $C/C($\sigma$)$ & $$\Delta_{11}$/${\sigma}$($\Delta_{11}$)$ & $P(${>}\Delta_1$)$ & $P$(>\chi^2)$$ & $C/C($\sigma$)$ & $$\Delta_{11}$/${\sigma}$($\Delta_{11}$)$ & $P(${>}\Delta_1$)$ \\
            \hline
            \noalign{\smallskip}
            total   & 0.666 &  +0.0   & -0.9  & 0.381 & 0.015  & -3.2   & -1.9   & 0.085 \\
            S       & 0.511 &  -0.7   & -1.2  & 0.434 & 0.225  & +0.4   & +0.8   & 0.383 \\
            SE      & 0.973 &  +0.1   & -0.9  & 0.569 & 0.031  & +2.0   & +2.8   & 0.015 \\
            SL      & 0.234 &  +0.5   & +0.8  & 0.209 & 0.460  & -0.1   & -0.5   & 0.345 \\
            SB      & 0.729 &  +0.3   & +1.0  & 0.454 & 0.285  & -1.0   & -0.2   & 0.497 \\
            SBE     & 0.739 &  +0.1   & -0.5  & 0.566 & 0.230  & -0.7   & +0.1   & 0.521 \\
            SBL     & 0.043 &  +1.8   & +1.7  & 0.046 & 0.620  & -0.9   & -0.2   & 0.872 \\
            RV1     & 0.910 &  +0.3   & +0.8  & 0.362 & 0.369  & -0.9   & -0.6   & 0.285 \\
            RV2     & 0.790 &  +0.3   & -1.0  & 0.496 & 0.925  & -0.4   & -0.2   & 0.887 \\
            RV3     & 0.050 &  +1.6   & -1.5  & 0.083 & 0.033  & -1.8   & -2.3   & 0.046 \\
            RV4     & 0.043 &  -2.3   & -1.5  & 0.116 & 0.636  & +0.2   & -0.7   & 0.692 \\
            Gr2     & 0.455 &  +0.6   & +0.8  & 0.861 & 0.033  & -1.8   & +1.7   & 0.116 \\
            Gr5     & 0.033 &  -1.4   & -2.0  & 0.085 & 0.516  & +0.4   & -0.4   & 0.548 \\
            \noalign{\smallskip}
            \hline
         \end{array}
     $$
\end{table*}
%-----------------------------------------------------------------

%----------------------------------------------------------------
\subsubsection{Morphology}

%----------------------------------------------------------------

In the spirals, the chi-square and auto correlation tests show
isotropy for both the trailing and leading modes. The first order
Fourier probability is found greater than 35\%, suggesting no
preferred alignment. However, the $\Delta_{11}$ value exceeds
1$\sigma$ limit (--1.2$\sigma$) in the trailing spirals. A hump at
90$^{\circ}$ is not enough to turn the
$\Delta_{11}$/$\sigma(\Delta_{11}$) $>$ 1.5 (Fig. 5a). Similarly,
a hump at 150$^{\circ}$ is not enough to make the
$\Delta_{11}$/$\sigma(\Delta_{11}$) $>$ 1.5 in the leading
spirals. Hence, the preferred alignment is not profounded in both
the leading and trailing spirals. Thus, we conclude a random
orientation of trailing and leading arm spirals.

% figure 10
\begin{figure} \vspace{0.0cm}
      \centering
      \includegraphics[height=4cm]{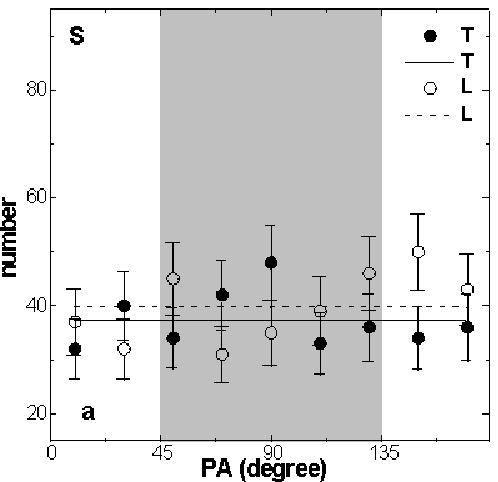}
      \includegraphics[height=4cm]{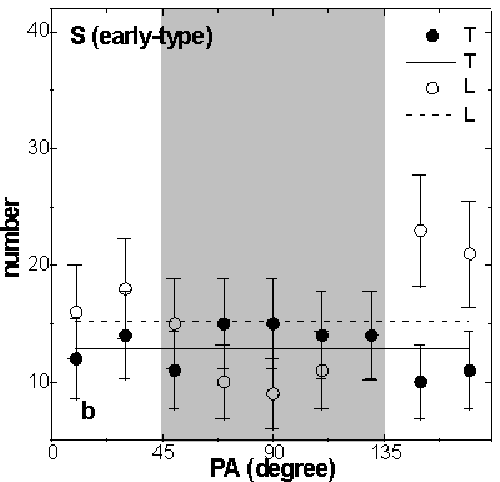}
      \includegraphics[height=4cm]{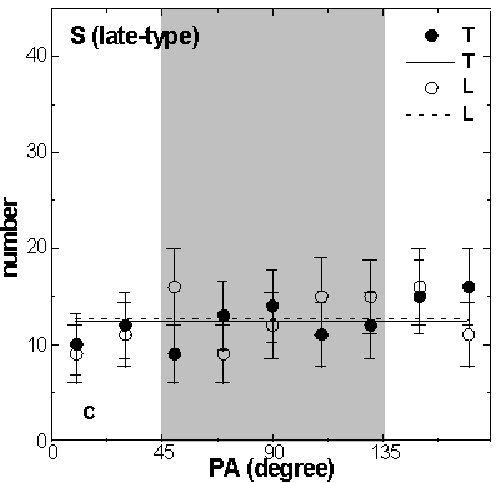}
      \includegraphics[height=4cm]{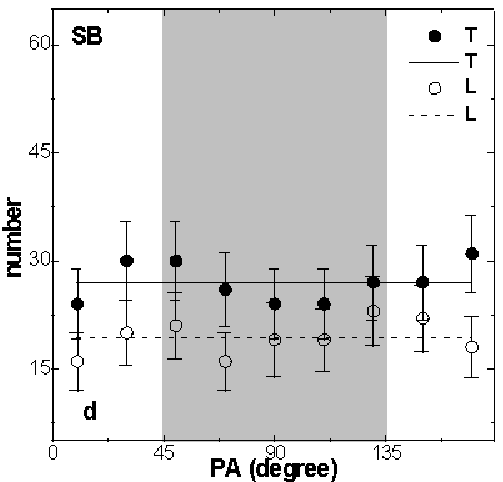}
      \includegraphics[height=4cm]{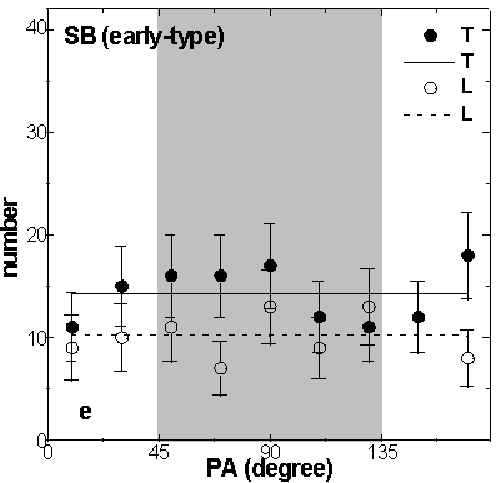}
      \includegraphics[height=4cm]{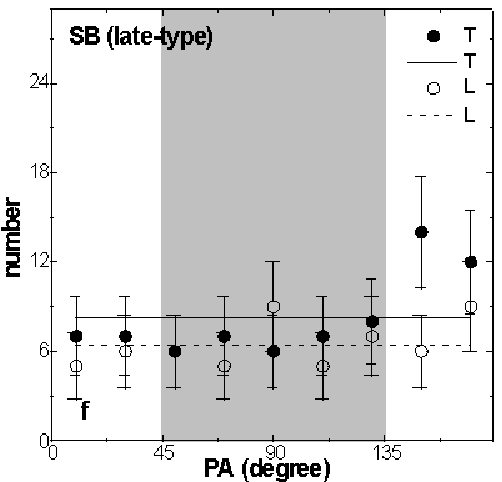}
      \caption[]{The equatorial PA-distribution of trailing and
      leading arm galaxies in the spirals (a), early-type spirals (b), late-type spirals (c), barred spirals (d),
      early-type barred spirals (e) and late-type barred spirals (f). The symbols, error
      bars, dashed lines and the explanations are analogous to Fig. 4.}
\end{figure}
%---------------------------------------------------------------

Early- and late-type trailing arm spirals show isotropy in all
three statistical tests (Table 2). No humps and the dips are seen
in the histograms (solid circles in Fig. 5b,c). Thus, the trailing
arm spirals show a random alignment in the PA-distribution. In the
subsample SE, all three statistical tests show anisotropy (Table
2). Two significant humps at $>$ 150$^{\circ}$ cause the first
order Fourier coefficient ($\Delta_{11}$) $>$ +2.5$\sigma$ (hollow
circle in Fig. 5b). Thus, a preferred alignment is noticed in the
early-type leading arm spirals: the galactic rotation axes tend to
lie in the equatorial plane. The late-type leading spirals show a
random alignment.

The spiral barred galaxies show a random alignment in both the
trailing and leading modes. In Fig. 5d, no deviation from the
expected distribution can be seen. All three statistical tests
support this result (Table 2). A similar result is found for the
early-type SB galaxies in both structural modes (Table 2, Fig.
5e).

The P$(>\chi^2)$ and P($>\Delta_1$) are found less than 5\%,
suggesting a preferred alignment for the late-type SB galaxies
having trailing arm (Table 2). The auto correlation coefficient
(C/C($\sigma$)) and the hump at $>$ 150$^{\circ}$ support this
result (Fig. 5f). The $\Delta_{11}$/$\sigma(\Delta_{11}$) is found
to be positive at 1.7$\sigma$ level, suggesting that the trailing
arm SBL galaxies tend to lie in the equatorial plane. Thus, the
late-type trailing and the leading arm SB galaxies show preferred
and random alignments, respectively.

%---------------------------------------------------------------
\subsubsection{Radial velocity}

The subsamples RV1 and RV2 show isotropy in all three statistical
tests (Table 2). No humps or dips can be seen in Figs. 6a,b. Thus,
the galaxies having radial velocity in the range 3\,000 km
s$^{-1}$ to 4\,000 km s$^{-1}$ show a random alignment for both
the leading and the trailing structural modes.

The humps at 90$^{\circ}$ ($>$2$\sigma$) and 110$^{\circ}$
($>$2$\sigma$) are found in the leading and trailing arm RV3
galaxies, respectively (Fig. 6c). These two significant humps lead
the subsample show anisotropy in the statistical tests (Table 2).
The $\Delta_{11}$ values are found negative  at $\geq$1.5 level,
suggesting a similar preferred alignment for both modes: the
galaxy rotation axes tend to be directed perpendicular to the
equatorial plane.

%----------------------------------------------------------------
% figure 10
\begin{figure} \vspace{0.0cm}
      \centering
      \includegraphics[height=4cm]{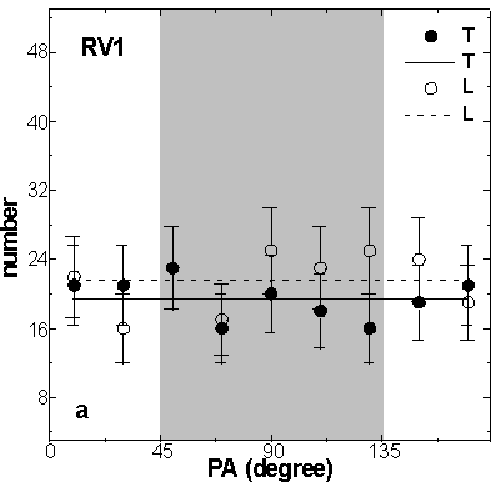}
      \includegraphics[height=4cm]{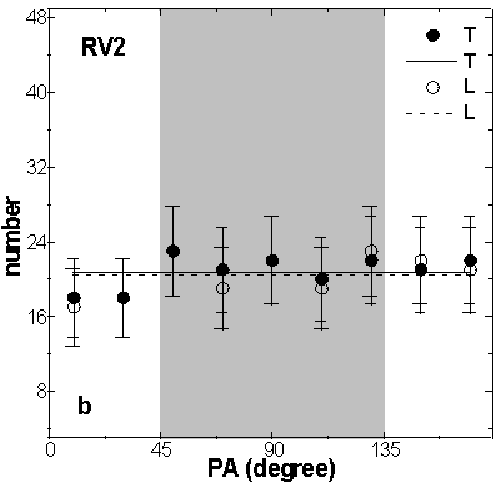}
      \includegraphics[height=4cm]{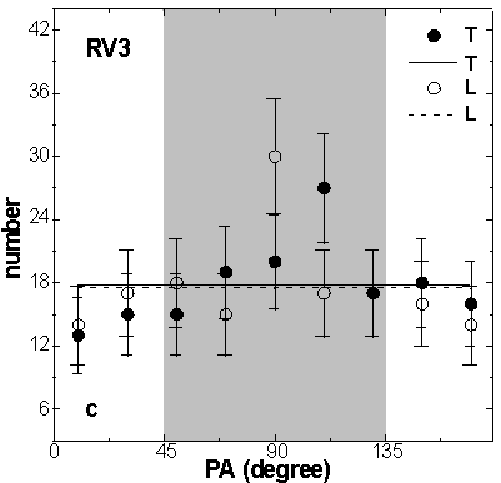}
      \includegraphics[height=4cm]{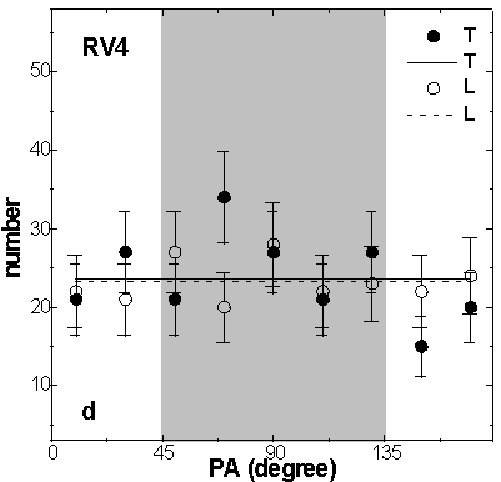}
      \caption[]{The equatorial PA-distribution of trailing and
      leading arm galaxies in RV1 (a), RV2 (b), RV3 (c) and RV4 (d). The abbreviations are listed
      in Table 1. The symbols, error bars, dashed lines and the explanations are analogous to Fig. 4.}
\end{figure}
%---------------------------------------------------------------

A hump at 70$^{\circ}$ ($>$1.5$\sigma$) and a dip at 150$^{\circ}$
($\sim$2$\sigma$) cause the trailing arm RV4 galaxies to show
anisotropy in all three statistical tests (Fig. 6d). Thus, the
trailing arm galaxies having radial velocity in the range 4\,500
km s$^{-1}$ to 5\,000 km s$^{-1}$ show a similar alignment as
shown by the subsample RV3: galactic planes of galaxies tend to
lie in the equatorial plane. The leading arm galaxies in the
subsample RV4 show a random alignment (Table 2, Fig. 6d).

%---------------------------------------------------------------
\subsubsection{Groups}

We do not study PA-distribution of leading and trailing arm
galaxies in the groups Gr1, Gr3, Gr4 and Gr6 because of poor
statistics (number $<$ 50). Fortunately, a very good correlation
between the leading and trailing arm galaxies are noticed in these
groups. In other words, we noticed that the chirality is not
violated in these groups.

We study the PA-distribution of leading and the trailing arm
galaxies in the groups Gr2 and Gr5, where the chiral symmetry seem
to be violated. In addition, the statistics is relatively better
in these two groups.
%----------------------------------------------------------------
% figure 10
\begin{figure} \vspace{0.0cm}
      \centering
      \includegraphics[height=4cm]{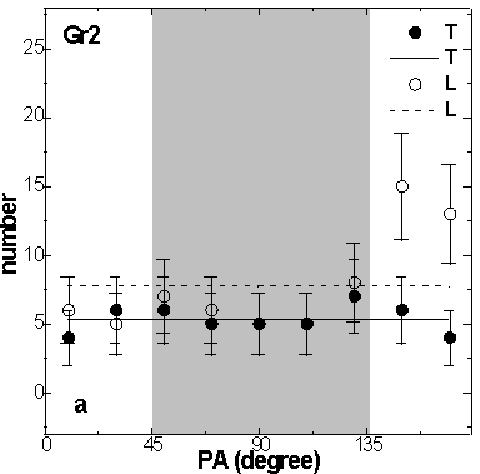}
      \includegraphics[height=4cm]{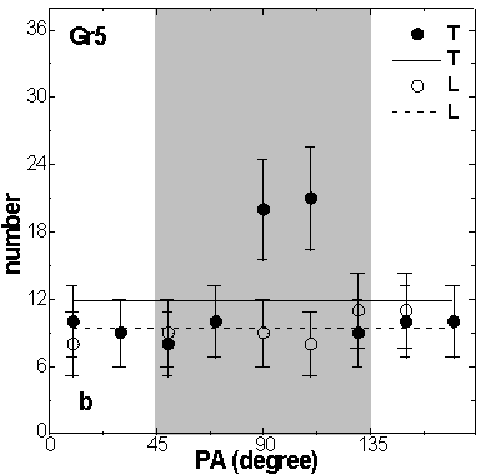}
      \caption[]{The equatorial PA-distribution of trailing and
      leading arm galaxies in the groups Gr2 and Gr5. The abbreviations are listed
      in Table 1. The symbols, error bars, dashed lines and the explanations are analogous to Fig. 4.}
\end{figure}
%---------------------------------------------------------------

In the group Gr2, leading arm galaxies dominate the trailing arm
galaxies. In this group, the leading arm galaxies show a preferred
alignment whereas trailing arm galaxies show a random alignment in
the PA-distribution. All three statistical tests suggest
anisotropy in the leading arm galaxies (Table 2). The humps at $>$
150$^{\circ}$ cause the $\Delta_{11}$ value to be positive at $>$
1.5$\sigma$ level (Fig. 7a), suggesting that the rotation axes of
leading arm galaxies in Gr2 tend to be oriented parallel the
equatorial plane.

The trailing arm galaxies dominate in the group Gr5.
Interestingly, a preferred alignment of trailing arm galaxies is
noticed in the PA-distribution. In Fig. 7b, two significant humps
at 90$^{\circ}$ ($\sim$2$\sigma$) and 110$^{\circ}$ ($>$2$\sigma$)
can be seen. These humps lead the subsample (trailing Gr5) to show
anisotropy in the statistical tests (Table 2). No preferred
alignment is noticed in the leading arm galaxies in this group.

Thus, the structural modes (leading or trailing) whose population
dominates in the groups show a preferred alignment in the
PA-distribution. This is an interesting result.

%---------------------------------------------------------------
\subsection{Discussion}
%--------------------------------------------------------------

Fig. 8a shows a comparison between the number ($\Delta$) and
position angle ($\Delta_{11}$/$\sigma(\Delta_{11}$)) distribution
of leading and trailing arm galaxies in the total sample and
subsamples. This plot deals the possible correlation between the
chirality (non-chirality) and the random (preferred) alignment in
the subsamples. The grey-shaded region represents the region of
isotropy and chirality for the $\Delta_{11}$/$\sigma(\Delta_{11}$)
and $\Delta$(\%), respectively.

Twenty five (out of 39, 64\%) subsamples lie in the grey-shaded
region (Fig. 8a), suggesting a good agreement between the chiral
property and the random alignment of the rotation axes of
galaxies. In four subsamples (SE, SBL, Gr2 and Gr5), a good
correlation between the preferred alignment and the achiral (i.e.,
non-chiral) property is noticed (Fig. 8a). Thus, It is found that
the random alignment of the PAs of galaxies lead the existence of
chiral property of galaxies.

%------------------------------------------------------------
\begin{figure}
\vspace{0.0cm}
      \centering
       \includegraphics[height=4.8cm]{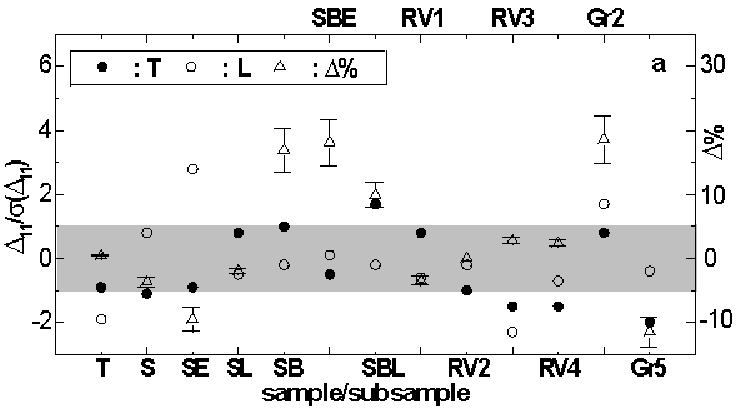}
      \includegraphics[height=5cm]{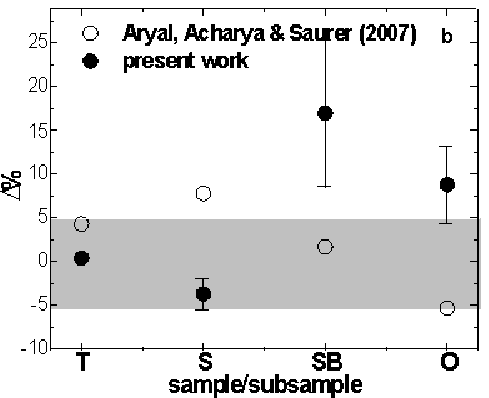}
      \caption[]{(a) A comparison between the number ($\Delta$\%) and
the position angle ($\Delta_{11}$/$\sigma(\Delta_{11}$))
distribution of leading and trailing arm galaxies in the total
sample and subsamples. (b) A comparison with the previous work
(Aryal, Acharya \& Saurer 2007). Error bars as in Fig. 2.}
\end{figure}
%--------------------------------------------------------------

Here we discuss our results with the results obtained by Aryal et
al. (2007). Aryal et al. (2007, hereafter Paper 1) compiled a
database of 667 leading and trailing arm galaxies in the LSC (RV
$<$ 3\,000 km s$^{-1}$) and studied the chiral property in the
total sample and 17 subsamples. In the present study, the database
is compiled from the field galaxies having RV in the range 3\,000
to 5\,000 km s$^{-1}$. Thus, we are moving deep inside the sky and
studying the existence of chiral property in the present work.

The distribution of the total trailing and leading structures in
the LSC (Paper 1) as well as in the field galaxies (present work)
are found homogeneous (Fig. 8b). The difference between the
leading and trailing arm galaxies are turned well within 5\% error
limit. This result indicate the fact that the chiral property is a
global phenomenon rather than a local phenomenon.

For the spirals, we noticed no deviation from the chirality.
However, Paper 1 found a slight deviation because of the presence
of the Virgo cluster. Interestingly, we noticed the violation of
chirality in the barred spirals (SB) whereas Paper 1 concluded the
SB galaxies as a chiral object. For other morphological types
(irregulars, morphologically unidentified galaxies), our result is
similar to that of the previous result.

In the PA-distribution, we noticed a random alignment for the
trailing arm galaxies whereas Paper 1 concludes no preferred
alignments for the leading arm galaxies. In this work, a preferred
alignment is noticed for the leading arm galaxies. In contrast to
this, a preferred alignment for the trailing arm galaxies is
concluded in Paper 1. This contradiction is interesting in the
sense that the nature of the databases are different in the past
(LSC galaxies having RV $<$ 3\,000 km s$^{-1}$) and the present
(field galaxies having 3\,000 $<$ RV (km s$^{-1}$) $\leq$ 5\,000)
study. The preferred alignment in both the studies are obviously
different. For leading arm LSC galaxies, they found that the
galactic rotation axes tend to lie in the equatorial plane. In our
case, we noticed that the rotation axes of the trailing arm
galaxies tend to be oriented perpendicular the equatorial plane.

Paper 1 concludes a similar preferred alignment in the leading and
trailing structures in the Virgo cluster galaxies: the galactic
rotation axes tend to lie in the equatorial plane. We noticed a
similar result for the early-type spirals (leading mode),
late-type barred spirals (trailing mode) and the group Gr2
(leading mode).

An important similarity is found in the past and present study:
late-type galaxies show chiral property, whereas this property is
found to be violated in the early-type galaxies.

In this way, we noticed few inconsistencies in the present and the
previous work. It should be remembered that these inconsistencies
are profounded either due to the poor statistics or because of the
bias in the sample classification.

We combined the database of the Paper 1 and the present work, and
studied the basis statistics. A strong chirality ($\Delta$ $<$
1\%, $\Delta(a\,sde)$ $<$ 0.010 and $\Delta(a)\%$ $<$ 5\%) is
found in the total and the spiral galaxies. However, the $\Delta$
value is found to be greater than 10\% for the barred spirals.
Thus, the total and the spiral galaxies having RVs less than
5\,000 km s$^{-1}$ strongly exhibit chiral symmetry. It seems that
the chirality loss sequence (spiral $\longrightarrow$ barred
spiral $\longrightarrow$ elliptical) as predicted by Capozziello
and Lattanzi (2006) might be true.

Aryal \& Saurer (2004, 2005b, 2006) and Aryal, Paudel \& Saurer
(2007) studied the spatial orientation of galaxies in 32 Abell
clusters of BM type I (2004), BM type III (2005b), BM type II-III
(2006) and BM type II (2007) and found a significant preferred
alignment in the late-type cluster (BM type II-III, BM type III).
They concluded that the randomness decreases systematically in
galaxy alignments from early-type (BM type I, II) to late-type (BM
type II-III, III) clusters. Thus, the existance of chirality in BM
type I cluster, as predicted by Capozziello and Lattanzi (2006)
might be true. Because we noticed a very good correlation between
the randomness and the chirality. Probably, this result reveals
the fact that the progressive loss of chirality might have some
connection with the rotationally supported (spirals, barred
spirals) to the randomized (lenticulars, ellipticals) system.
Thus, we suspect that the dynamical processes in the cluster
evolution (such as late-type clusters) give rise to a dynamical
loss of chirality. In other word, a good correlation between the
achirality and anisotropy can be suspected for the late-type
clusters. It would be interesting to test this prediction by
analysing the chiral property of spirals in the late-type clusters
in the future.

%_______________________________________________________________
\section{Conclusion}

We studied the chiral property of 1\,621 field galaxies having
radial velocity (RV) in the range 3\,000 km s$^{-1}$ to 5\,000 km
s$^{-1}$. The distribution of leading and trailing structural
modes is studied in the total sample and 34 subsamples. To examine
non-random effects, the equatorial position angle (PA)
distribution of galaxies in the total sample and subsamples are
studied. In order to discriminate anisotropy from the isotropy we
have performed three statistical tests: chi-square,
auto-correlation and the Fourier. Our results are as follows:

\begin{enumerate}

\item The homogeneous distribution of the total trailing and the
total leading arm galaxies is found, suggesting the existence of
chiral symmetry in the field galaxies having RVs 3\,000 km
s$^{-1}$ to 5\,000 km s$^{-1}$. The PA-distribution of trailing
arm galaxies is found to be random, whereas preferred alignment is
noticed for leading arm galaxies. It is found that the galactic
rotation axes of leading structural modes tend to be oriented
perpendicular the equatorial plane.

\item Leading structural modes are found 3.7\% $\pm$ 1.8\% more
than that of the trailing modes in the spirals whereas a
significant dominance (17\% $\pm$ 8.5\%) of trailing modes are
noticed in the barred spirals. This difference is found $>$ 8\%
for the irregulars and the morphologically unidentified galaxies.
A random alignment is noticed in the PA-distribution of leading
and trailing spirals. Thus, it is noticed that the random
alignment of the PAs of galaxies lead the existence of chiral
property of galaxies. In other words, a good correlation between
the preferred alignment and the chiral symmetry breaking is found.
This result verifies the previous result (Paper 1).

\item The chirality is found stronger for the late-type spirals
(Sc, Scd, Sd and Sm) than that of early-type (Sa, Sab, Sb and
Sbc). Similar result is found for the late-type barred spirals.
Thus, the late-type galaxies are the best candidates for the
chiral objects than that of the early-types.

\item A very good correlation between the number of leading and
trailing arm galaxies are found in the RV subsamples. All 4
subsamples show the $\Delta$ value less than 5\%. Thus, we
conclude that the chirality of field galaxies remain invariant
with the global expansion.

\item The galaxies having RVs 3\,000 km s$^{-1}$ to 4\,000 km
s$^{-1}$ show a random alignment for both the leading and the
trailing structural modes. The rotation axes of leading and
trailing arm galaxies having 4\,000 $<$ RV (km s$^{-1}$) $\leq$
4\,500 tend to be oriented perpendicular the equatorial plane.

\item The galaxies possesses the chiral symmetry in $\sim$ 80\%
area of the sky. This property is found to be violated in few
groups of galaxies. Two such groups (Gr2 and Gr8) are identified.
In these groups, the structural dominance and the preferred
alignments of galaxies are found to oppose each other.

\end{enumerate}

Aryal et al. (2007) studied the chiral property of the Local
Supercluster galaxies (RV $<$ 3\,000 km s$^{-1}$). If we include
their database, we found a strong ($\Delta$ $<$ 1\%) chiral
behavior by the total and the spiral galaxies. However, the
$\Delta$ value is found to be greater than 10\% for the barred
spirals. Hence, the total and the spiral galaxies having RV $<$
5\,000 km s$^{-1}$ show chiral property. Thus, the chirality-loss
sequence (spiral $\longrightarrow$ barred spiral $\longrightarrow$
elliptical) as proposed by Capozziello and Lattanzi (2006) might
be true.

The true structural mode of a galaxy must involve a determination
of which side of the galaxy is closer to the observer (Binney and
Tremaine 1987). Three-dimensional determination of the leading and
the trailing arm patterns in the galaxies is a very important
problem. We intend to address this problem in the future.

%______________________________________________________________

\section*{Acknowledgments}
This research has made use of the NASA/IPAC Extragalactic Database
(NED) which is operated by the Jet Propulsion Laboratory,
California Institute of Technology, under contract with the
National Aeronautics and Space Administration. We acknowledge
Prof. Udayraj Khanal and Prof. Mukunda Mani Aryal for insightful
discussions. One author (RP) acknowledge Central Department of
Physics, Tribhuvan University, Kirtipur, for providing various
forms of support for their masters thesis.


\begin{thebibliography}{3}
\bibitem{Abell} Abell, G.O., Corwin, H.G., Olowin, R.P.: {\it Astrophys. J. Supp.} {\bf 70}, 1 (1989)
\bibitem{Aryal6} Aryal, B., Paudel, S., Saurer, W.: {\it Astronom. Astrophys.} {\bf 479}, 397 (2008)
\bibitem{Aryal1} Aryal, B., Acharya, S., Saurer, W.: {\it Astrophys. Space Sci.}, Bib. Co. tmp.4A, in press (2007)(Paper 1)
\bibitem{Aryal4} Aryal, B., Saurer, W.: {\it Monthly Notices Royal Astron. Soc.} {\bf 366}, 438 (2006)
\bibitem{Aryal2} Aryal, B., Saurer, W.: {\it Astronom. Astrophys.} {\bf 432}, 841 (2005a)
\bibitem{Aryal2} Aryal, B., Saurer, W.: {\it Astronom. Astrophys.} {\bf 432}, 431 (2005b)
\bibitem{Aryal1} Aryal, B., Saurer, W.: {\it Astronom. Astrophys.} {\bf 425}, 871 (2004)
\bibitem{Aryal3} Aryal, B., Saurer, W.: {\it Astronom. Astrophys. lett.} {\bf 364}, L97 (2000)
\bibitem{Bagchi} Bagchi, M. et al.: {\it Astronom. Astrophys.} {\bf 450}, 431 (2006)
\bibitem{Binney} Binney J., Tremaine, S.: {\it Galactic Dynamics}, Princeton Univ. press, Princeton, New Jersey (1987)
\bibitem{Brun} Brunzendorf, J., Meusinger H.: {\it Astronom. Astrophys. Supp. Ser.} {\bf 139}, 141 (1999)
\bibitem{CapLat1} Capozziello, S., Lattanzi, A.: {\it Astrophys. Space Sci.} {\bf 301}, 1-4, 189 (2006)
\bibitem{Chang} Chang, L. et al.: {\it Phy. Rev. C.}, {\bf 75}, id015201 (2007)
\bibitem{Corwin} Corwin, H.G., de Vaucouleurs, A., de Vaucouleurs, G.: {\it Univ. Texas Monogr. Astron.} {\bf 4}, 1 (1985)
\bibitem{deVauc} de Vaucouleurs, G., de Vaucouleurs, A., Corwin, et al.: {\it Third Reference Catalogue of Bright Galaxies}, Springer-Verlag, New York (1991)
\bibitem{Fall} Fall, S.M.: in: {\it Progress in Cosmology}, Proc. of the Oxford Intl. Symp., Oxford, Dordrecht, D. Reidel Publishing Co., 347-356 (1982)
\bibitem{Garcia} Garcia-Garcia, A.M., Cuevas, E.: {\it Phy. Rev. B.}, {\bf 74}, id113101 (2006)
\bibitem{Godlow1} Godlowski, W.: {\it Monthly Notices Royal Astron. Soc.} {\bf 265}, 874 (1993)
\bibitem{Godlow2} Godlowski, W.: {\it Monthly Notices Royal Astron. Soc.} {\bf 271}, 19 (1994)
\bibitem{Kodaira} Kodaira, K., Okamura, S., Ichikawa, S.: {\it Photometric Atlas of Northern Bright Galaxies}, Univ. of Tokyo Press, Tokyo (1990)
\bibitem{Laubert} Lauberts, A.: {\it ESO/Uppsala Survey of the ESO B Atlas}, Garching bei Muenchen (1982)
\bibitem{Liddle} Liddle, A.R., Lyth, D.H.: {\it Cosmological Inflation and Large-Scale Structure}, Cambridge Univ. Press, Cambridge (2000)
\bibitem{Nilson1} Nilson, P.: {\it Uppsala General Catalogue of Galaxies}, Nova Acta Uppsala University, Ser. V:A, Vol.1 (1973)
\bibitem{Nilson2} Nilson, P.: {\it Upps. Astron. Obs. Rep.}, 5 (1974)
\bibitem{Oort1} Oort, J.H.: {\it Science}, {\bf 170}, 1363 (1970a)
\bibitem{Oort2} Oort, J.H.: {\it Astronom. Astrophys.}, {\bf 7}, 405 (1970b)
\bibitem{Peacock} Peacock, J.A.: {\it Cosmological Physics}, Cambridge Univ. Press, Cambridge (1999)
\bibitem{Pasha} {\bf Pasha, I.I.: {\it Sov. Astron. Lett.}, {\bf 11}, 1 (1985)}
\bibitem{Shectman} Shectman, S.A., Landy, S.D., Oemler, A. et al.: 1996, {\it Astrophys. J.} {\bf 470}, 172 (1996)
\bibitem{Struble} Struble, M.F., Rodd, H.J.: {\it Astrophys. J. Supp.} {\bf 125}, 355 (1999)
\bibitem{Sugai} {\bf Sugai, H. \& Iye, M.: {\it Monthly Notices Royal Astron. Soc.} {\bf 276}, 327 (1995)}
\bibitem{Thom} {\bf Thomasson, M. et al.:  {\it Astronom. Astrophys.}, {\bf 211}, 25 (1989)}
\end{thebibliography}
\end{document}